\documentclass{rspublic}
\usepackage[final]{graphics}
\begin{document}

\title[Noise Filtering]{Application of Wavelets to Filtering of Noisy Data}

\author[U. Pen]{Ue-Li Pen}

\affiliation{Canadian Institute for Theoretical Astrophysics,
University of Toronto, 60 St. George St., Toronto}

\label{firstpage}

\maketitle

\begin{abstract}{wavelets, optimal filtering, wiener filtering}
I discuss approaches to optimally remove noise from images.
A generalization of Wiener filtering to Non-Gaussian distributions
and wavelets is described, as well as an approach to measure the
errors in the reconstructed images.  We argue that the wavelet basis
is highly advantageous over either Fourier or real space analysis if
the data is intermittent in nature, i.e. if the filling factor of
objects is small.
\end{abstract}

\section{Introduction}

In astronomy, the collection of data is often limited by the presence
of background noise.  Various methods are used to filter the noise
while retaining as much ``useful'' information as possible.  In recent
years, wavelets have played an increasing role in astrophysical data
analysis.  It provides for a general parameter-free procedure to look
for objects of varying size scales.  In the case of the Cosmic
Microwave Background (CMB) one is interested in the non-Gaussian
component in the presence of Gaussian noise and signal.  An
application of wavelets is presented by Tenorio et al (1999).  This
paper generalizes their analysis beyond the thresholding
approximation.  X-ray images are also frequently noise dominated,
caused by instrumental and cosmic background.  Successful wavelet
reconstructions were achieved by Damiani et al (1997a,b).

At times generic tests for non-Gaussianity are desired.  Inflationary
theories predict, for example, that the intrinsic fluctuations in the
CMB are Gaussian, while topological defect theories predict
non-Gaussianity.  A full test for non-Gaussianity requires measuring
all N-point distributions, which is computationally not tractable for
realistic CMB maps.  Hobson et al (1998) have shown that wavelets are
a more sensitive discriminant between cosmic string and inflationary
theories if one examines only the one point distribution function of
basis coefficients.

For Gaussian random processes, Fourier modes are statistically
independent.   Current theories of structure formation start from
an initially linear Gaussian random field which grows non-linear
through gravitational instability.  Non-linearity occurs through
processes local in real space.  Wavelets provide a natural basis which
compromise between locality in real and Fourier space.  Pando \& Fang
(1996) have applied the wavelet decomposition in this spirit to the
high redshift $L_\alpha$ systems which are in the transition from
linear to non-linear regimes, and are thus well analyzed by the wavelet
decomposition. 

We will concentrate in this paper on the specific case of data layed
out on a two dimensional grid, where each grid point is called a {\it
pixel}.  Such images are typically obtained through various imaging
instruments, including CCD arrays on optical telescopes,
photomultiplier arrays on X-ray telescopes, differential radiometry
measurements using bolometers in the radio band, etc.  In many
instances, the images are dominated by noise.  In the optical, the sky
noise from atmospheric scatter, zodiacal light, and extragalactic
backgrounds, sets a constant flux background to any observation.  CCD
detectors essentially count photons, and are limited by the Poissonian
discreteness of their arrival.  A deep exposure is dominated by sky
background, which is subtracted from the image to obtain the features
and objects of interest.  Since the intensity of the sky noise is
constant, it has a Poissonian error with standard deviation $e\propto
n^{1/2}$, where $n$ is the photon count per pixel.  After subtracting
the sky average, this fluctuating component remains as white noise in
the image.  For large modern telescopes, images are exposed to near
the CCD saturation limit, with typical values of $n\sim 10^4$.  The
Poisson noise is well described by Gaussian statistics in this limit.

We would like to pose the problem of filtering out as much of the
noise as possible, while maximally retaining the data.  In certain
instances, optimal methods are possible.  If we know the data to
consist of astronomical point objects, which have a shape on the grid
given by the atmospheric spreading or telescope optics, we can test
the likelihood at each pixel that a point source was centered there.
The iterative application of this procedure is implemented in the
routine {\it clean} of the Astronomical Image Processing Software
(AIPS) (Cornwell \& Braun 1989).

If the sources are not point-like, or the atmospheric point spread
function varies significantly across the field, {\it clean} is no
longer optimal.  In this paper we examine an approach to implement a
generic noise filter using a wavelet basis.  In section
\ref{sec:classical} we first review two popular filtering techniques,
thresholding and Wiener.  In section \ref{sec:opt} we generalize
Wiener filtering to inherit the advantages of thresholding.  A
Bayesian approach to image reconstruction (Vidakovic 1998) is used,
where we use the data itself to estimate the prior distribution of
wavelet coefficients.  We recover Wiener filtering for Gaussian data.
Some concrete examples are shown in section \ref{sec:example}.

\section{Classical Filters}
\label{sec:classical}

\subsection{Thresholding}

A common approach to supressing noise is known as thresholding.  If
the amplitude of the noise is known, one picks a specific threshold
value, for example $\tau=3\sigma_{\rm noise}$ to set a cutoff at three
times the standard deviation of the noise.  All pixels less than this
threshold are set to zero.  This approach is useful if we wish to
minimize false detections, and if all sources of signal occupy only a
single pixel.  It is clearly not optimal for extended sources, but
often used due to its simplicity.  The basic shortcoming is its
neglect of correlated signals which covers many pixels.  The choice of
threshold also needs to be determined heuristically.  We will attempt
to quantify this procedure.

\subsection{Wiener Filtering}

\begin{figure}
\resizebox{\textwidth}{!}{
\includegraphics{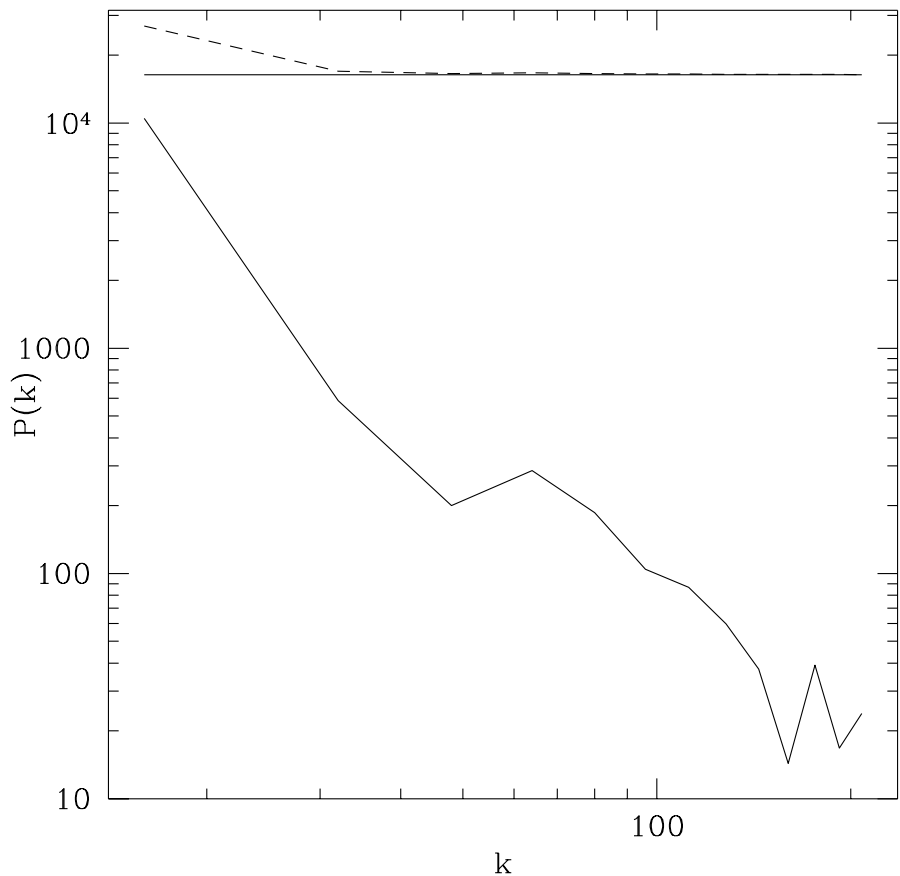}
}
\caption{The power spectrum of figure \protect{\ref{plate:org}}.  The
dashed line is the power spectrum measured from the noisy data.  The
horizontal line is the noise.  The lower solid line is the difference
between the measured spectrum and the noise.  We see that the
measurement of the difference becomes noise limited at large $k$.
}
\label{fig:pkwien}
\end{figure}

In the specific case that both the signal and the noise are Gaussian
random fields, an optimal filter can be constructed which minimizes
the impact of the noise.  If the noise and signal are stationary
Gaussian processes, Fourier space is the optimal basis where all modes
are uncorrelated.  In other geometries, one needs to expand in
signal-to-noise eigenmodes (see e.g. Vogeley and Szalay 1996).  One
needs to know both the power spectrum of the data, and the power
spectrum of the noise.  We use the least square norm as a measure of
goodness of reconstruction.  Let $E$ be the reconstructed image, $U$
the original image and $N$ the noise.  The noisy image is called
$D=U+N$.  We want to minimize the error $e=\langle (E-U)^2\rangle$.
For a linear process, $E=\alpha D$.  For our stationary Gaussian
random field, different Fourier modes are independent, and the optimal
solution is $\alpha=\langle U^2\rangle/\langle D^2\rangle$.  $U^2$ is
the intrinsic power spectrum.  Usually, $D^2$ can be estimated from
the data, and if the noise power spectrum is known, the difference can
be estimated subject to measurement scatter as shown in figure
\ref{fig:pkwien}.  Often, the powerspectrum decays with increasing wave
number (decreasing length scale): $\langle U^2\rangle =c k^{-n}$.  For
white noise with unit variance, we then obtain $\alpha= c/(k^n+c)$,
which tends to one for small $k$ and zero for large $k$.  We really
only need to know the parameters $c,n$ in the crossover region
$E^2\sim 1$.  In section \ref{sec:example} we will illustrate a worked
example.

Wiener filtering is very different from thresholding, since modes are
scaled by a constant factor independent of the actual amplitude of the
mode.  If a particular mode is an outlier far above the noise,
the algorithm would still force it to be scaled back.  This can
clearly be disadvantageous for highly non-Gaussian distributions.  If
the data is localized in space, but sparse, the Fourier modes dilute
the signal into the noise, thus reducing signal significantly as is
seen in the examples in section \ref{sec:example}.  Furthermore,
choosing $\alpha$ independent of $D$ is only optimal for Gaussian
distributions.   One can generalize as follows:

\section{Non-Gaussian Filtering}
\label{sec:opt}

We can extend Wiener filtering to Non-Gaussian Probability Density
Functions (PDFs) if the PDF is
known and the modes are still statistically independent.  We will
denote the PDF for a given mode as $\Theta(u)$ which 
describes a random variable $U$.  When Gaussian white noise with unit variance
${\cal N}(0,1)$ is added, we obtain a new random variable $D=U+{\cal N}(0,1)$
with PDF 
$f(d)=(2\pi)^{-1/2}\int \Theta(u) \exp(-(u-d)^2/2) du$.  We can 
calculate the conditional probability $P(U|D)=P(D|U)P(U)/P(D)$ using
Bayes' theorem.  For the posterior conditional expectation value we obtain
\begin{eqnarray}
\langle U|D=d \rangle &=& \frac{1}{\sqrt{2\pi}f(d)}
  \int \exp[-(u-d)^2/2] \Theta(u) u du \nonumber\\
&=& D +  \frac{1}{\sqrt{2\pi}f(d)}
  \partial_d \int \exp[-(u-d)^2/2] \Theta(u) du \nonumber \\
&=& D+ (\ln f)'(d).
\label{eqn:bayes}
\end{eqnarray}
Similarly, we can calculate the posterior variance 
\begin{equation}
 \left\langle (U-\bar{U})^2 | D=d \right\rangle = 1+(\ln f)''(d).
\label{eqn:var}
\end{equation}
For a Gaussian prior with variance $\sigma$, equation
(\ref{eqn:bayes}) reduces to Wiener filtering.  We have a generalized
form for $\alpha=1+(\ln f)'/D$.  For distributions with
long tails, $(\ln f)' \sim 0$, $\alpha \sim 1$, and we leave the
outliers alone, just as thresholding would 
suggest. 

For real data, we have two challenges:

1. estimating the prior distribution $\Theta$.

2. finding a basis in which $\Theta$ is most non-Gaussian.

\subsection{Estimating Prior $\Theta$}

The general non-Gaussian PDF on a grid is a function of $N$ variables,
where $N$ is the number of pixels.  It is generally not possible to
obtain a complete description of this large dimensional space
(D. Field, this proceedings).  It is often possible, however, to make
simplifying assumptions.  We consider two descriptions: Fourier space
and wavelet space.  We will assume that the one point distributions of
modes are non-Gaussian, but that they are still statistically
independent.  In that case, one only needs to specify the PDF for each
mode.  In a hierarchical basis, where different basis functions sample
characteristic length scales, we further assume a scaling form of the
prior PDF $\Theta_l(u)=l^{-\beta}\Theta(u/l^\beta)$.  Here $l\sim 1/k$
is the characteristic length scale, for example the inverse wave
number in the case of Fourier modes.
For images, we often have $\beta
\sim 1$.  

Wavelets still have a characteristic scale, and we can similarly
assume scaling of the PDF.  In analogy with Wiener filtering, we first
determine the scale dependence.  For computational simplicity, we use
Cartesian product wavelets (Meyer 1992).  Each basis function has two
scales, call them $2^i,2^j$.  The real space support of each wavelet
has area $A\propto 2^{i+j}$, and we find empirically that the variance
depends strongly on that area.  The scaling relation does not directly
apply for $i\ne j$, and we introduce a lowest order correction using
$\ln(\sigma) = c_1 (i+j)+c_2(i-j)^2+c_3$.  We then determine the best
fit parameters $c_i$ from the data.  The actual PDF may depend on the
length scale $i+j$ and the elongation $i-j$ of the wavelet basis.  One
could parameterize the PDF, and solve for this dependence (Vidakovic
1998), or bin all scales together to measure a non-parametric scale
averaged PDF.  We will pursue the latter.

\begin{figure}
\resizebox{\textwidth}{!}{
\includegraphics{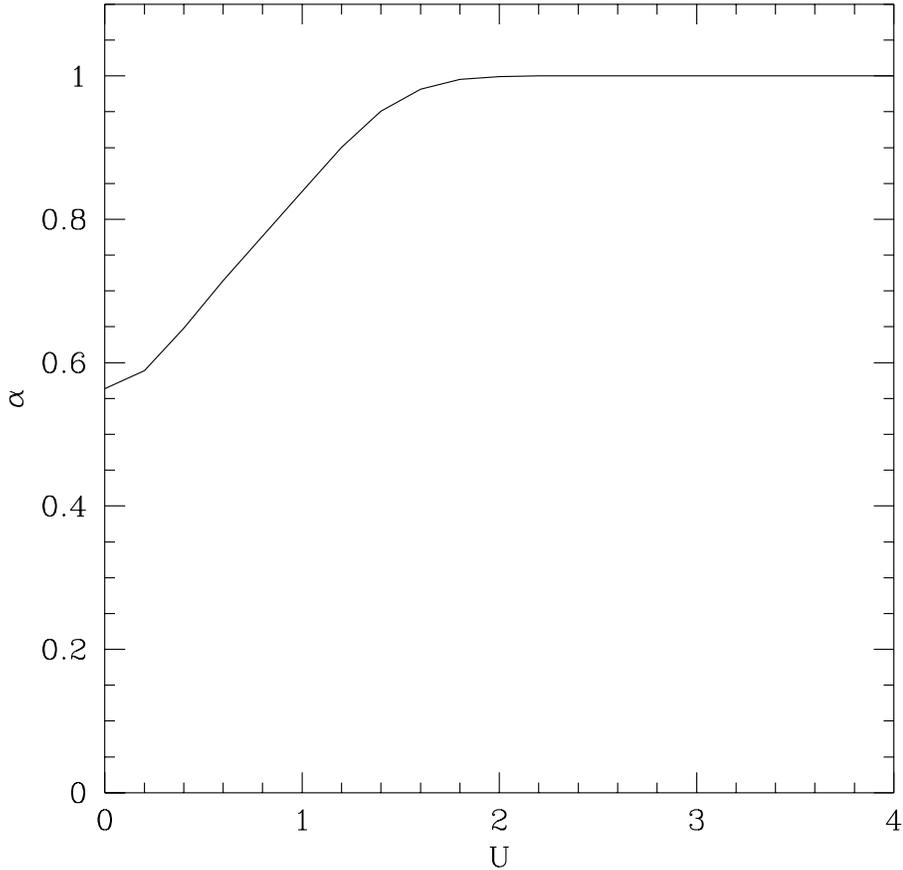}
}
\caption{The optimal filter function $\alpha$ for the non-Gaussian
wavelet model at $\sigma_{\rm noise}=\sigma_{\rm data}$.  $U$ is given
in units of the total standard deviation $\sigma_{\rm
abs}^2=\sigma_{\rm noise}+\sigma_{\rm data}$. 
At small
amplitudes, it is similar to a Wiener filter $\alpha=1/2$, but limits
to 1 for large outliers.}
\label{fig:alph}
\end{figure}

The observed variance is the
intrinsic variance of $\Theta$ plus the noise variance of ${\cal N}$, so
the variance $\sigma_{\rm
intrinsic}^2=\sigma_{\rm obs}^2-\sigma_{\rm noise}^2$ has error
$\propto \sigma_{\rm obs^2}/n$ where $n$ is the number of coefficients
at the same length scale.  We weigh the data accordingly.
Because most wavelet modes are at short scales, most of the weight
will come near the noise threshold, which is what we desire.  We now
proceed to estimate $f(d)$.  Our Ansatz now assumes
$\Theta_{ij}(u)\propto\Theta(u/\exp[c_1(i+j)+c_2(i-j)^2+c_3])$ where
$\Theta(u)$ has unit variance.  We can only directly measure $f_{ij}$.
We sort these in descending order of variance, $f_m$.  Again,
typically the largest scale modes will have the largest variance.  In
the images explored here, we find typical values of $c_1$ between 1
and 2, while $c_2 \sim -0.2$.
For the largest variance modes, noise is least important.  From the
data, we directly estimate a binned PDF for the largest scale modes.  
By hypothesis,  $D=U/l^\beta+{\cal N}(0,\sigma_{l,\rm
noise})$. 
We  reduce the larger scale PDF
by convolving it with the difference of noise levels to
obtain an initial guess for the smaller scale PDF:
\begin{equation}
f'_{l'}(d) = \frac{(l'/l)^\beta}{\sqrt{\pi}}
\int_{-\infty}^\infty f_l\left[u (l/l')^\beta\right]
\exp\left[-\frac{(u-d)^2}{2(\sigma_{l',\rm noise}^2-\sigma_{l,\rm noise}^2) }
\right] du.
\end{equation}
To this we add the actual histogram of wavelets coefficients at the
smaller scale.
We continue this hierarchy to obtain an increasingly better estimate
of the PDF, having used the information from each scale. 
In figure \ref{fig:alph} we show the optimal weighting function
obtained for the examples in section \ref{sec:example}.  

On the largest scales, the PDF will be poorly defined because
relatively few wavelets lie in that regime.  The current
implementation performs no filtering, i.e. sets $\alpha=1$ for the
largest scales.  A potential improvement could be implemented: Within
the scaling hypothesis, we can deconvolve the noisy $f(D)$ obtained
from small scales to estimate the PDF on large scales.  The errors in
the PDF estimation are themselves Poissonian, and in the limit that we
have many points per PDF bin, we can treat those as Gaussian.  The
deconvolution can then be optimally filtered to maximize the use of
the large number of small scale wavelets to infer the PDF of large
scale wavelets.  Of course, the non-Gaussian wavelet analysis could
then be recursively applied to estimate the PDF.  Instead of the
Bayesian prior PDF, we would then specify a prior for the prior.  This
possibility will be explored in future work.

\subsection{Maximizing Non-Gaussianity using Wavelets}
\label{subsec:nongauss}

Errors are smallest if a large number of coefficients are near zero,
and when the modes are close to being statistically independent.  Let
us consider several extreme cases and their optimal strategies.
Imagine that we have an image consisting of true uncorrelated point
sources, and each point source only occupies one pixel.  Further
assume that only a very small fraction $\epsilon$ of possible pixels
are occupied, but when a point source is present, it has a constant
luminosity $L$.  And then add a uniform white noise background with
unit variance.  In Fourier space, each mode has unit variance from the
noise, and variance $L^2 \epsilon$ from the point sources.  We easily
see that it will be impossible to distinguish signal from noise if
$L^2 \epsilon < 1$.  In real space, white noise is also uncorrelated,
so we are justified to treat each pixel separately.  Now we can easily
distinguish signal from noise if $L > \sqrt{-\ln(\epsilon)}$.  If
$L=10$ and $\epsilon=0.001$, we have a situation where the signal is
easy to detect in real space and difficult in Fourier space, and in
fact the optimal filter (\ref{eqn:bayes}) is optimal in real space
where the points are statistically independent.  In Fourier space,
even though the covariance between modes is zero, modes are not
independent. 

Now consider the more realistic case that objects occupy more than one
pixel, but are still localized in space, and only have a small
covering fraction.  This is the case of intermittent information.  The
optimal basis will depend on the actual shape of the objects, but it
is clear that we want basis functions which are localized.  Wavelets
are a very general basis to achieve this, which sample objects of any
size scale, and are able to effectively excise large empty regions.
We expect PDFs to be more strongly non-Gaussian in wavelet space than
either real or Fourier space.

In this formulation, we obtain not only a filtered image, but also an
estimate of the residual noise, and a noise map.  For each wavelet
coefficient we find its posterior variance using (\ref{eqn:var}).  The
inverse wavelet transform then constructs a noise variance map on the
image grid.

\section{Examples}
\label{sec:example}

\begin{figure}
\resizebox{\textwidth}{!}{
\includegraphics{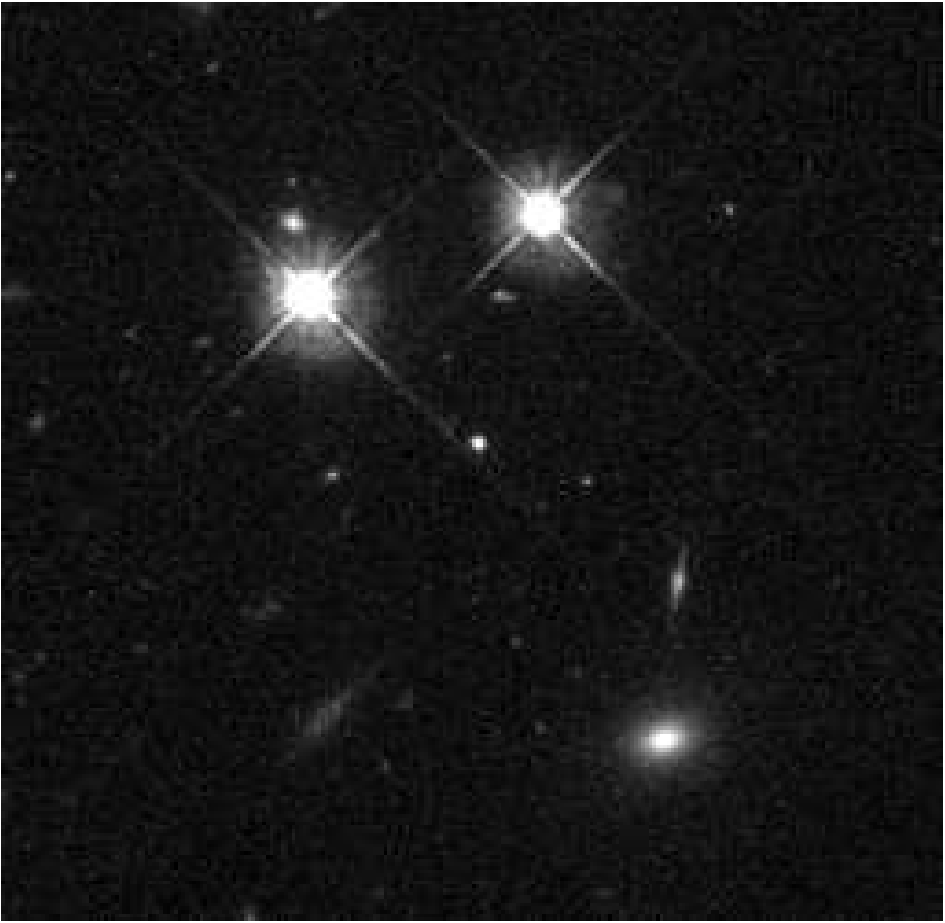}
}
\caption{The original image, taken from the Space Telescope web page
{\tt www.stsci.edu}.  It is the 100,000th image taken with HST for
C. Steidel of Caltech.}
\label{plate:org}
\end{figure}

\begin{figure}
\resizebox{\textwidth}{!}{
\includegraphics{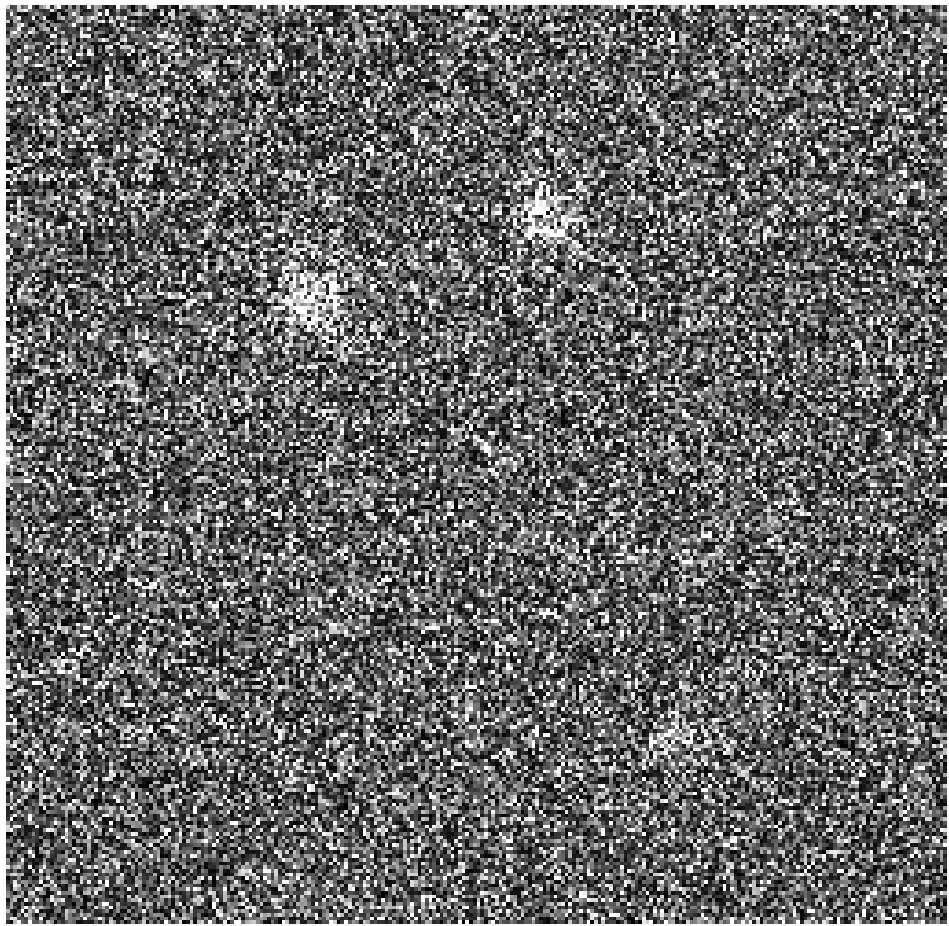}
}
\caption{Figure \protect\ref{plate:org} with substantial noise added.}
\label{plate:noise}
\end{figure}

In order to be able to compare the performance of the filtering
algorithm, we use as example an image to which the noise is added by
hand.  The de-noised result can then be compared to the 'truth'.
We have taken a random image from the Hubble Space telescope,
in this case the 100,000th image (PI: C. Steidel).  The original
picture is shown in figure \ref{plate:org}.  The gray scale is from 0
to 255.  At the top are two bright stars with the telescope support
structure diffraction spikes clearly showing.  The extended objects
are galaxies.  We then add noise with variance 128, which is shown in
figure \ref{plate:noise}.  The mean signal to noise ratio of the image
is 1/4.  We can tell by eye that a small number of regions still
protrude from the noise.

The power spectrum of the noisy image is shown in figure
\ref{fig:pkwien}.  We use the known expectation value of the noise
variance.  The subtraction of the noise can be performed even when the
noise substantially dominates over the signal, as can be seen in the
image.  In most astronomical applications, noise is instrumentally
induced and the distribution of the noise is very well documented.
Blank field exposures, for example, often provide an empirical
measurement. 

\begin{figure}
\resizebox{\textwidth}{!}{
\includegraphics{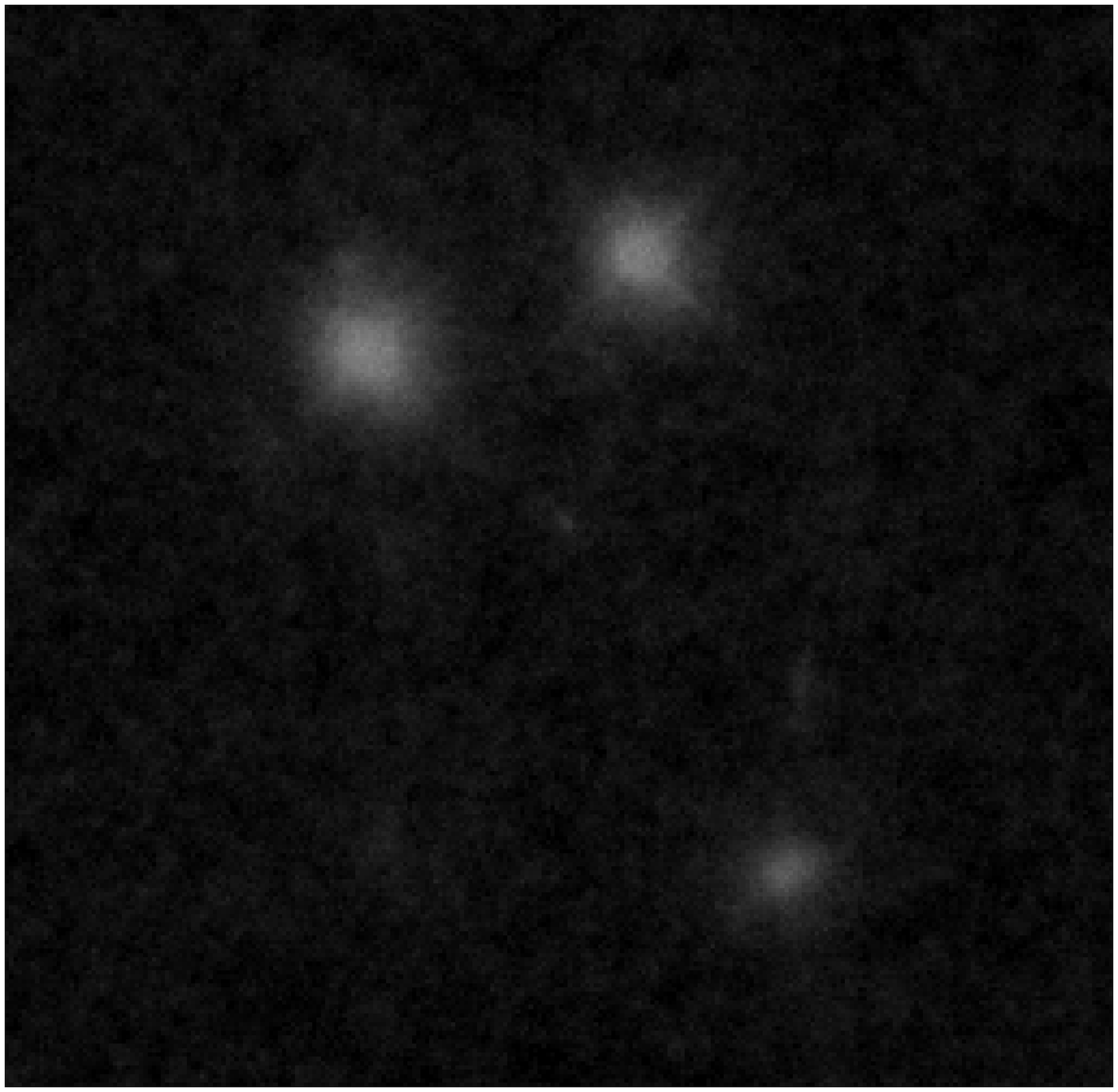}
}
\caption{Wiener filtered.  We see that the linear scaling function
substantially reduces the flux in the bright stars.}
\label{plate:wiener}
\end{figure}

We first apply a Wiener filter, with the result shown in figure
\ref{plate:wiener}.  We notice immediately the key feature: all
amplitudes are scaled by the noise, so the bright stars have been down
scaled significantly.  The noise on the star was less than unity, but
each Fourier mode gets contributions from the star as well as the
global noise of the image.  The situation worsens if the filling
factor of the signal regions is small.  The mean intensity of the
image is stored in the $k=0$ mode, which is not significantly affected
by noise.  While total flux is approximately conserved, the flux on
each of the objects is non-locally scattered over the whole image by
the Wiener filter process.

\begin{figure}
\resizebox{\textwidth}{!}{
\includegraphics{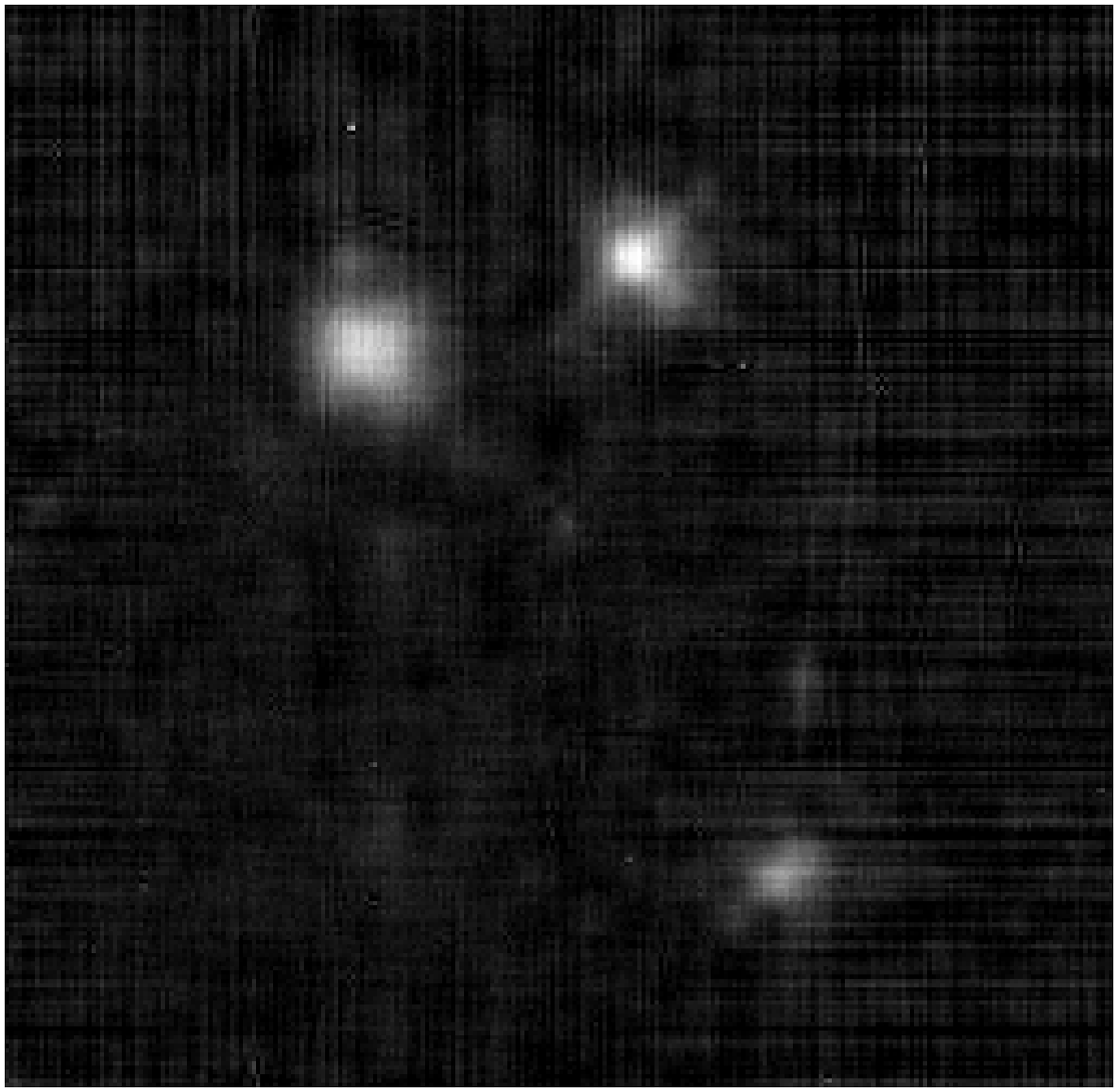}
}
\caption{Non-Gaussian wavelet filtered.  
Several of the features that had been lost in the Wiener filtering
process are recovered here.}
\label{plate:dwt}
\end{figure}

The optimal Bayesian wavelet filter is shown in figure \ref{plate:dwt}.
A Daubechies 12 wavelet was used, and the prior PDF reconstructed
using the scaling assumption described in section   \ref{sec:opt}.
We see immediately that the amplitudes on the bright objects are much
more accurate.  We see also that the faint vertical edge-on spiral on
the lower right just above the bright elliptical is clearly visible in
this image, while it had almost disappeared  in the Wiener filter.

\begin{figure}
\resizebox{\textwidth}{!}{
\includegraphics{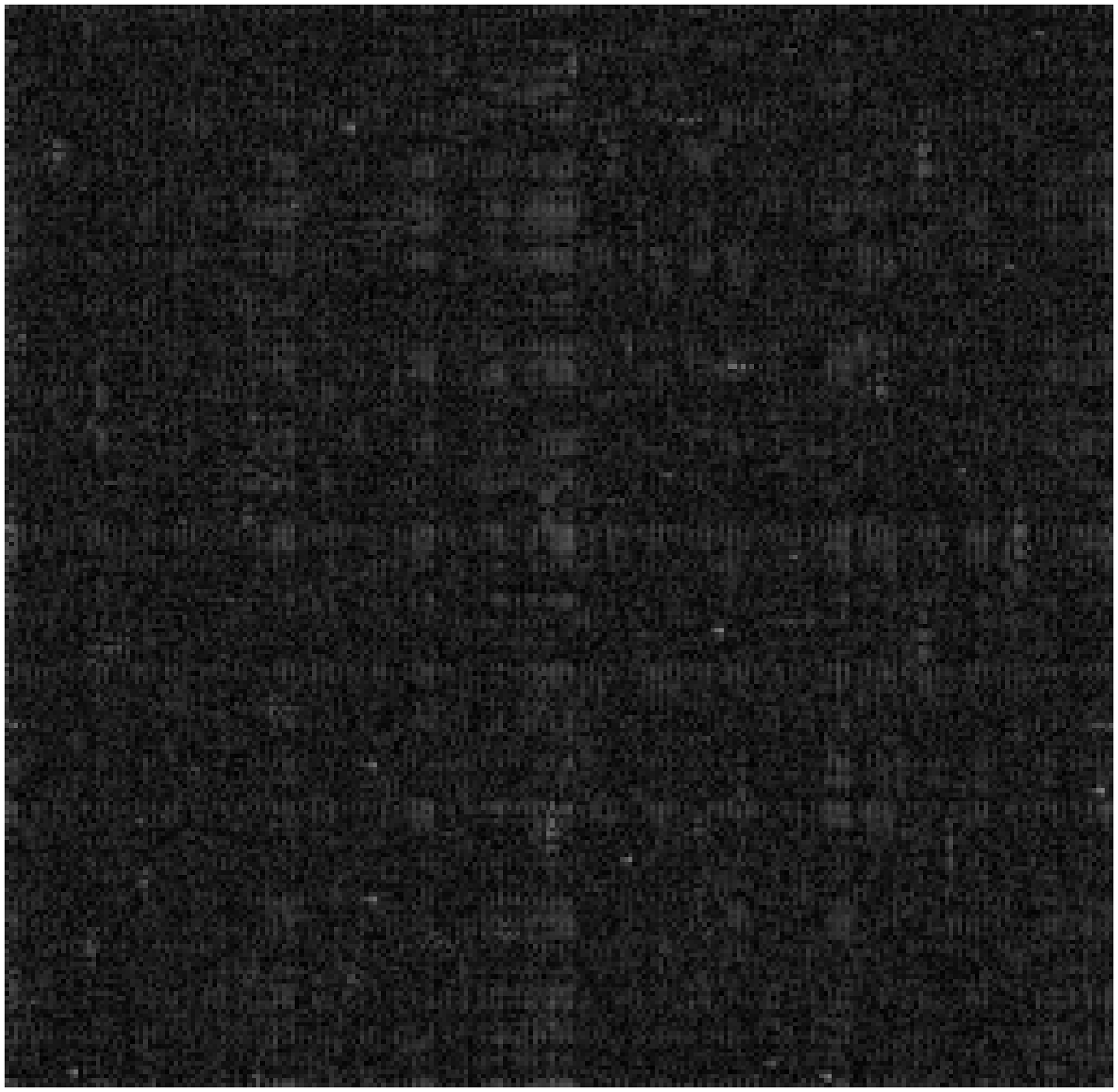}
}
\caption{Error map.  Plotted is the posterior Bayesian variance.
We see that some features, for example the small
dot in the upper left, have large errors associated with them, and are
therefore artefacts.}
\label{plate:err}
\end{figure}

The Bayesian approach allows us to estimate the error in the
reconstruction using Equation (\ref{eqn:var}).  We show the result in
figure \ref{plate:err}.  We can immediately see that some features in
the reconstructed map, for example the second faint dot above the
bright star on the upper left, have large errors associated with them,
and are indeed artefacts of reconstruction.  Additionally, certain
wavelets experience large random errors.  These appear as checkered
'wavelet' patterns on both the reconstructed image and the error map.

\section{Discussion}

Fourier space has the advantage that for translation invariant
processes, different modes are pairwise uncorrelated.  If modes were
truly independent, the optimal filter for each $k$ mode would also be
globally optimal.  As we have seen from the example in section
\ref{sec:opt}\ref{subsec:nongauss}, processes which are local in real
space are not optimally processed in Fourier space, since different
Fourier modes are not independent.  Wavelet modes are not independent,
either.  For typical data, the correlations are relatively sparse.  In
the astronomical images under consideration, the stars and galaxies
are relatively uncorrelated with each other.  Wavelets with compact
support sample a limited region in space, and wavelets which do not
overlap on the same objects on the grid will be close to independent.
Even for Gaussian random fields, wavelets are close to optimal since
they are relatively local in Fourier space.  Their overlap in Fourier
space leads to residual correlations which are neglected.  We see that
wavelets are typically close to optimal, even though they are never
truly optimal.  But in the absence of a full prior, they allow us to
work with generic data sets and usually outperform Wiener filtering.

In our analysis, we have used Cartesian product Daubechies wavelets.
These are preferentially aligned along the grid axes.  In the wavelet
filtered map (figure \ref{plate:dwt}) we see residuals aligned with the
coordinate axes.  Recent work by Kingsbury (this proceedings) using
complex wavelets would probably alleviate this problem.  The complex
wavelets have a factor of two redundancy, which is used in part to
sample spatial translations and rotational directions more
homogeneously and isotropically.

%One person's signal is often another person's noise.  A current
%direction of active research is the precision measurement of the
%cosmic microwave background (CMB).  Inflationary theories predict Gaussian
%fluctuations with nearly scale invariant power spectra.  Recently the
%COBE satellite has measured the CMB to the exquisite precision of a
%few $\mu$ K, as shown in figure \ref{plate:cobe}.  For the analysis of
%the background fluctuations, the galactic plane is usually excised, as
%shown in the bottom panel.  Using the wavelet analysis described
%above, it may be possible to filter out the cosmic signal to obtain a
%map of galactic emission.  This can then be subtracted from the all
%sky map, to measure the Gaussian contribution behind the galaxy.
%Future work will explore these directions.
%
%\begin{figure}
%\caption{The all sky CMB maps measured with the COBE satellite.  The
%image is in galactic coordinates, and we can clearly see the emission
%by dust and plasma in the galaxy along the equator.  With optimal
%filtering, one may be able to subtract the cosmic fluctuations to
%obtain a full galactic map.}
%\label{plate:cobe}
%\end{figure}

\section{Conclusions}

We have presented a generalized noise filtering algorithm.  Using the
Ansatz that the PDF of mode or pixel coefficients is scale invariant,
we can use the observed data set to estimate the PDF.  By application
of Bayes' theorem, we reconstruct the filter map and noise map.  The
noise map gives us an estimate of the error, which tells us the
performance of the particular basis used and the confidence level of
each reconstructed feature.  Based on comparison with controlled data,
we find that the error estimates typically overestimate the true error
by about a factor of two.

We argued that wavelet bases are advantageous for data with a small
duty cycle that is localized in real space.  This covers a large class
of astronomical images, and images where the salient information is
intermittently present.

\begin{acknowledgements}
I would like to thank Iain Johnstone, David Donoho and Robert
Crittenden for helpful discussions.  I am most grateful to the Bernard
Silvermann and the Royal Society for organizing this discussion meeting.
\end{acknowledgements}

\end{document}